\documentclass[%
 reprint,
 amsmath,amssymb,
 aps,
 prl,
]{revtex4-2}

\usepackage{graphicx}
\usepackage{dcolumn}
\usepackage{bm}
\usepackage{xspace}
\usepackage{enumitem} 

\newcommand{\tia}{\textsc{type i}a\xspace}
\newcommand{\tib}{\textsc{type i}b\xspace}
\newcommand{\tii}{\textsc{type ii}\xspace}

\begin{document}

\title{Seismic Background Limitation of Lunar Gravitational-wave Detectors}

\author{Jan Harms}
\email{jan.harms@gssi.it}
\affiliation{Gran Sasso Science Institute (GSSI), I-67100 L'Aquila, Italy}
\affiliation{INFN, Laboratori Nazionali del Gran Sasso, I-67100 Assergi, Italy}

\date{\today}

\begin{abstract}
New concepts were recently proposed for gravitational-wave (GW) detectors on the Moon. These include laser-interferometric detectors, proposed as free-range or optical-fiber interferometers, and inertial acceleration sensors. Some of them exploit the response of the Moon to GWs, others follow the design of current laser-interferometric GW detectors, which directly measure the gravitational strain with suspended optics.  All of these ideas emerged since the Moon offers an extremely quiet geophysical environment compared to Earth, but at the same time, one must realize that even the quiet lunar environment sets limits to the sensitivity of lunar GW detectors. In this article, we compare the proposed mission concepts in terms of their response to GWs and evaluate how they are affected by the lunar seismic background. We discuss available mitigation strategies. From these analyses, we infer the prime observation band of each detector concept.
\end{abstract}

\maketitle

We are at the beginning of a new era of lunar exploration, which also brings opportunities for revolutionary scientific experiments on the Moon. One of the oldest ideas for a lunar fundamental-physics experiment is the deployment of a gravitational-wave (GW) detector. An instrument called the Lunar Surface Gravimeter was developed under the coordination of Joseph Weber and deployed on the Moon with Apollo 17 in 1972 \cite{GiEA1977}. It implemented a proof mass as inertial reference to measure lunar surface vibrations caused by GWs. The experiment did not reach design performance, but this did not cause direct harm to the field of GW detection since even at best performance, we know today that the sensitivity would not have been enough to observe GWs. 

In the past few years, improved lunar GW detector concepts were proposed. The Lunar Gravitational-wave Antenna (LGWA) adopts the basic idea of Weber to use an inertial acceleration measurement to detect GWs. LGWA implements a method to reduce the impact of the lunar seismic background on the GW observations, and offers an improved technological design to meet the sensitivity requirements \cite{HaEA2021a}. The Lunar Seismic and Gravitational Antenna (LSGA) uses a laser-interferometric measurement of the seismic strain of the Moon caused by GWs \cite{KaEA2020}. We call these two concepts \textsc{type i}, since they exploit the response of the Moon to GWs. We refer to the inertial acceleration measurement as \tia and the seismic strain measurement as \tib. The Gravitational-Wave Lunar Observatory for Cosmology (GLOC) \cite{JaLo2021} and soon after the Laser Interferometer on the Moon (LION) \cite{AmEA2021} were proposed as laser-interferometric GW detectors with suspended test masses much like the current terrestrial GW detectors LIGO, Virgo, KAGRA \cite{LSC2015,AcEA2015,AkEA2018}. We call these concepts \tii since they attempt to decouple the GW measurement from ground motion.

In this paper, we will not make any detailed reference to the technologies of the proposed concepts, but instead, we provide a comparison between the concepts in terms of their GW response and susceptibility to the seismic background. For this purpose, we divide the seismic background into two components. The first part consists of the strong disturbances produced by larger meteoroid impacts and shallow or deep moonquakes. The rate and magnitude of such events is greatly reduced compared to Earth \cite{NaEA1981}. Still, they will reduce the duty cycle of \textsc{type i} detectors like strong ground motion reduces the duty cycle of terrestrial detectors today \cite{MuEA2019a}. Such events could also produce excess noise in \tii detectors, but reduced by the seismic-isolation system. Instead, since moonquakes are generally much weaker than terrestrial events (by a few orders of magnitude in amplitude), we can use our experience with terrestrial GW detectors to conclude that operation of \tii detectors, which generally have a more complex optomechanical response to external forces, will likely not be interrupted by lunar seismic events. Interruption here refers to the control of the suspended optics and the laser beam. 

More important for an assessment of seismic noise in lunar GW detector data is the spectrum $\delta x_\text{seis}(f)$ of continuous surface vibrations produced at any time by the overlap of a large number of weak seismic waves, e.g., from impacts of small meteoroids and from tails of moonquake phases. An upper limit of about 10$^{-10}$\,m/Hz$^{1/2}$ between 0.1\,Hz -- 1\,Hz was obtained from Apollo seismometer data \cite{CoHa2014c}, corresponding to the instrument noise of the lunar seismometers. A modeled estimate of the seismic background was provided by Lognonné et al \cite{LoEA2009}. Their results set an upper limit to the spectral density of surface displacement at 1\,Hz of about $2\cdot 10^{-14}$\,m/Hz$^{1/2}$ assuming that the full root-mean square of their simulated background is collected uniformly in the decihertz band. The predicted lunar background is up 6 orders of magnitude weaker in amplitude than the terrestrial background (depending on frequency). Nonetheless, it seems likely that lunar GW detectors will either be background limited in the decihertz band (\textsc{type i}), or require seismic isolation (\tii). Below the decihertz band, where the normal-mode formalism provides an effective description of the seismic field, the background was estimated to be negligible for LGWA and probably is negligible for all \textsc{type i} concepts \cite{HaEA2021a}. 

\paragraph{Detector response} There are two commonly used coordinate systems to calculate the response of elastic bodies to GWs: a local Lorentz (LL) frame, or using the transverse-traceless (TT) gauge \cite{ThBl2017}. Gravitational strain $h$ couples to the mass distribution of the elastic body in a LL frame, while in TT gauge, it couples to gradients of the shear modulus \cite{Dys1969,Ben1983}. For \tia and \tii detectors, the detector response must also consider the effect of the GW on the motion of the proof or test masses, which depends as well on the choice of the coordinate system. A freely falling test mass does not experience a coordinate change in TT gauge.

We start with a simple gedankenexperiment. Let us imagine a liquid planet. On the planet floats a boat with an inertial accelerometer to measure vibrations of the planet's surface. We also establish two laser links to a second boat at distance $L$ to be able to measure (a) the planet's strain, and (b) to measure gravitational strain using a pair of suspended optics forming a Fabry-Perot resonator. Now, adopting the LL point of view, when a GW passes, its quadrupolar tidal field will set fluid cells into motion. The liquid planet does not have any stiffness to counteract the gravitational strain (its shear modulus is equal to zero), which means that all fluid cells follow the field of the GW. Now, what do our three detectors measure? The \tia detector does not measure anything since its proof mass follows the motion of the planet's surface. There is no differential signal between the surface and the proof mass. The \tib and \tii detectors both observe a strain signal $0.5L\cdot h$, which corresponds to the distance change between the two boats or two suspended test masses. 

For the following discussion, we need to introduce the separation of the frequency band into a low-frequency band and a high-frequency band. The low-frequency band extends from the resonance frequency of the lowest-order quadrupole mode of the Moon (expected to be close to 1\,mHz) to the resonance frequency $f_n$ of the quadrupole normal mode $n$ where the product $Q_nL_n$ starts to be smaller than the radius $R_\text{moon}=1740\,$km of the Moon. Here, $Q_n$ is the quality factor of mode $n$, while $L_n$ is the effective GW interaction length of mode $n$. In other words, the frequency where $Q_nL_n=R_\text{moon}$ marks the transition from elastic to inertial response. We do not know precisely where in frequency this transition happens, but it is expected to happen around $f_\text{tr} =20\,$mHz \cite{HaEA2021a}. From here on, we focus on the high-frequency band.

For a solid body like the Moon, its mass distribution cannot follow the GW strain field. However, we know that at sufficiently high frequencies, it almost does so. Dyson calculated the high-frequency limit of the GW response of a homogeneous body as measured by an inertial accelerometer, and he found that it goes as $h\lambda/(2\pi)$ for the horizontal displacement signal, where $\lambda$ is the length of seismic shear waves. This means that towards high frequencies, the response of a uniform body decreases with $1/f$. Considering a LL frame with its origin at the Moon's center, this means that at high frequencies, the signal from the proof-mass displacement $0.5h R_\text{moon}$ is almost perfectly compensated by the inertial response of the Moon to GWs. In other words, with respect to its GW response, the Moon behaves almost like a liquid at high frequencies. Accordingly, if we take Dyson's simplified high-frequency approximation for granted, it follows that \tib detectors see a signal $0.5hL$ above $f_\text{tr}$, which is equal to the signal of \tii detectors. We can summarize the results in terms of effective detector baselines:
\begin{equation}
\begin{split}
    \mathcal B_{\text{\tia}} &= Q_\text{eff}(f)L_\text{eff}(f),\\
    \mathcal B_{\text{\tib}} &= L,\\
    \mathcal B_{\text{\tii}} &= L.
\end{split}
\end{equation}
Here, $Q_\text{eff}(f)$, $L_\text{eff}(f)$ are effective quality factors and interaction lengths of the Moon, so that we can take into account that the Moon is not a homogeneous body, and the overall GW response is a sum over all quadrupole normal modes. These baselines are approximately valid at $f>f_\text{tr}$ for \tia and \tib detectors. For \tii detectors, note that it has the properties of a \tib detector well below the lowest-frequency suspension resonances, but the coupling of a GW with \tib and \tii detectors is anyway the same. 

\paragraph{Seismic background} The GW response can now be confronted with the noise produced by the seismic background to assess the corresponding sensitivity limitation. We focus on the decihertz band since this is where sensitivity limitations of the seismic background are expected. For the \tib and \tii detectors, we assume that $L$ is at least 10\,km, e.g., one could imagine a deployment of interferometer stations at the rim of the Shackleton crater. In this case, throughout the decihertz band, correlations of seismic displacements between the ends of the interferometers would be small enough to ignore the common-mode rejection of the seismic noise in GW measurements. We can then immediately write down the background-limited sensitivity curves:
\begin{equation}
\begin{split}
    h_\text{\text{\tia}}&=2\delta x_\text{seis}(f)/(Q_\text{eff}(f)L_\text{eff}(f)),\\
    h_\text{\text{\tib}}&=2\delta x_\text{seis}(f)/L\\
    h_\text{\text{\tii}}&=2\mathcal S_\text{iso}(f)\delta x_\text{seis}(f)/L,
\end{split} 
\end{equation}
where $S_\text{iso}(f)$ is the attenuation function achieved by the seismic-isolation system. Let us start with the simplest case, which are the \tib detectors. Assuming a seismic background of $10^{-14}$\,m/Hz$^{1/2}$ at 1\,Hz and a baseline of $L=10$\,km of \tib detectors, we find a background limited strain sensitivity of $2\cdot 10^{-18}$\,Hz$^{-1/2}$, which is not good enough for GW science in this band. The seismic background would have to be rejected by at least two orders of magnitude (see below for background suppression methods). Another way to intuitively understand the strong limitation by the seismic background in \tib detectors is that seismic displacement produced by local events changes over much smaller distances than GW-induced seismic displacement, since the latter always takes the form of a quadrupole mode.

Concerning the \tia detectors, the important question is if the product $Q_\text{eff}(f)L_\text{eff}$ is sufficiently large in the decihertz band. If the Moon had a homogeneous geology, then the baseline at 1\,Hz would be about 1\,km, which is shorter than conceivable baselines of \tib detectors. However, we have reasons to believe that the lunar geology is favorable for GW detection. It was observed that moonquakes have a very long decay time with coda quality factors of meteoroid impacts going up to a several 1000 in the decihertz band \cite{NuEA2020}. We think that effective baselines of $1000$\,km are conceivable at 1\,Hz (and longer baselines at lower frequencies). This leads to a background limited sensitivity of \tia detectors of about $2\cdot 10^{-20}$\,Hz$^{-1/2}$. This would be good enough for GW science, and we will argue below that the sensitivity can be significantly improved with background rejection techniques. Note that the signal $Q_\text{eff}(f)L_\text{eff}$ also appears in \tib detectors as strain $Q_\text{eff}(f)L_\text{eff}\cdot 2/R_\text{moon}\ll 1$ in the decihertz band, which makes it negligible compared to the inertial strain signal.

Finally, with the previous estimate of the \tib sensitivity limit in the decihertz band, we conclude that the attenuation factor provided by the seismic-isolation system of \tii detectors needs to be at least about $S_\text{iso}(f)\sim 10^{-2}$ for GW observations. In order to reach the GLOC sensitivity target at 1\,Hz \cite{JaLo2021}, $S_\text{iso}(1\,\text{Hz})\sim 10^{-5}$ would be necessary.

\paragraph{Background mitigation} Four methods were proposed to reduce the impact of the seismic background on the measurement: 
\begin{itemize}
    \item Increase the number of readouts and obtain an advantage by averaging noise (\tia, \tib);
    \item Perform an optimized coherent noise cancellation using a sensor array to analyze the seismic field (\tia, \tib, \tii);
    \item Install a seismic-isolation system (\tii);
    \item Site selection (\tia, \tib, \tii).
\end{itemize}
The first method is straight-forward to achieve with \tia detectors by deploying multiple accelerometers as suggested for LGWA. Instead of doing a simple averaging of data, it is favorable to perform a coherent noise cancellation with array configurations optimized using surrogate models \cite{BaEA2020}. The noise-cancellation filters can be so-called Wiener filters, but also time-varying noise-cancellation filters are possible, e.g., using Kalman filters or neural networks. The achievable noise reduction needs to be assessed in field tests, e.g., by deploying sensor arrays in volcanic landscapes, which serve as analog of the lunar regolith. 

In order to achieve distributed strain sensing for background reduction in \tib detectors, it was proposed to deploy optical fibers as additional strainmeters. However, these require similar sensitivity to the main \tib detector to be effective for background reduction, which means that fundamental thermal-noise limitations must be overcome \cite{DoEA2016}. Some background reduction could also be achieved with accelerometers as used for \tia detectors. Due to the significantly higher susceptibility to the seismic background, we believe that the \tib detector is not a good concept for decihertz GW observations. 

Seismic isolation for \tii detectors is above all a practical challenge of installation and commissioning. With some of the proposed future seismic-isolation systems for terrestrial detectors already showing (mild) attenuation in the decihertz band \cite{ET2020}, and taking into account the strongly reduced gravitational acceleration on the Moon ($g\approx 1.62\,$m/s$^2$), there does not seem to be a fundamental show stopper. However, one needs to carefully analyze such a system in terms of noise introduced by the suspension and control, and above all, one needs to propose a way to install such a system on the lunar surface. Current and proposed isolation systems for terrestrial detectors are very massive structures that require an almost continuous on-site presence of people to commission them and maintain operation. \tia sensors could play a crucial role for \tii detectors as part of an active seismic-isolation system \cite{MaEA2014}, or to perform a feed-forward cancellation of residual seismic noise in data of \tii detectors \cite{DrEA2019}. 

Finally, when it comes to the seismic background, site selection is mostly a question about whether to deploy in a sunlit part of the lunar surface or in a permanently shadowed region (PSR) at the lunar poles \cite{GlEA2014,KlEA2021,HaEA2021b}. The PSRs have greatly reduced and stable surface temperatures, which is expected to greatly reduce ground motion from low-magnitude thermal moonquakes, which were observed in great number with the Apollo seismometers \cite{DuSu1974}. It is also well-known that changes in irradiation lead to local, thermal responses of the ground or of the deployed payload \cite{StEA2020}. The impact of these effects on \textsc{type i} detectors must be analyzed carefully as part of a site-selection process.

\paragraph{Observation bands} 
\begin{figure*}
\includegraphics[width=0.9\textwidth]{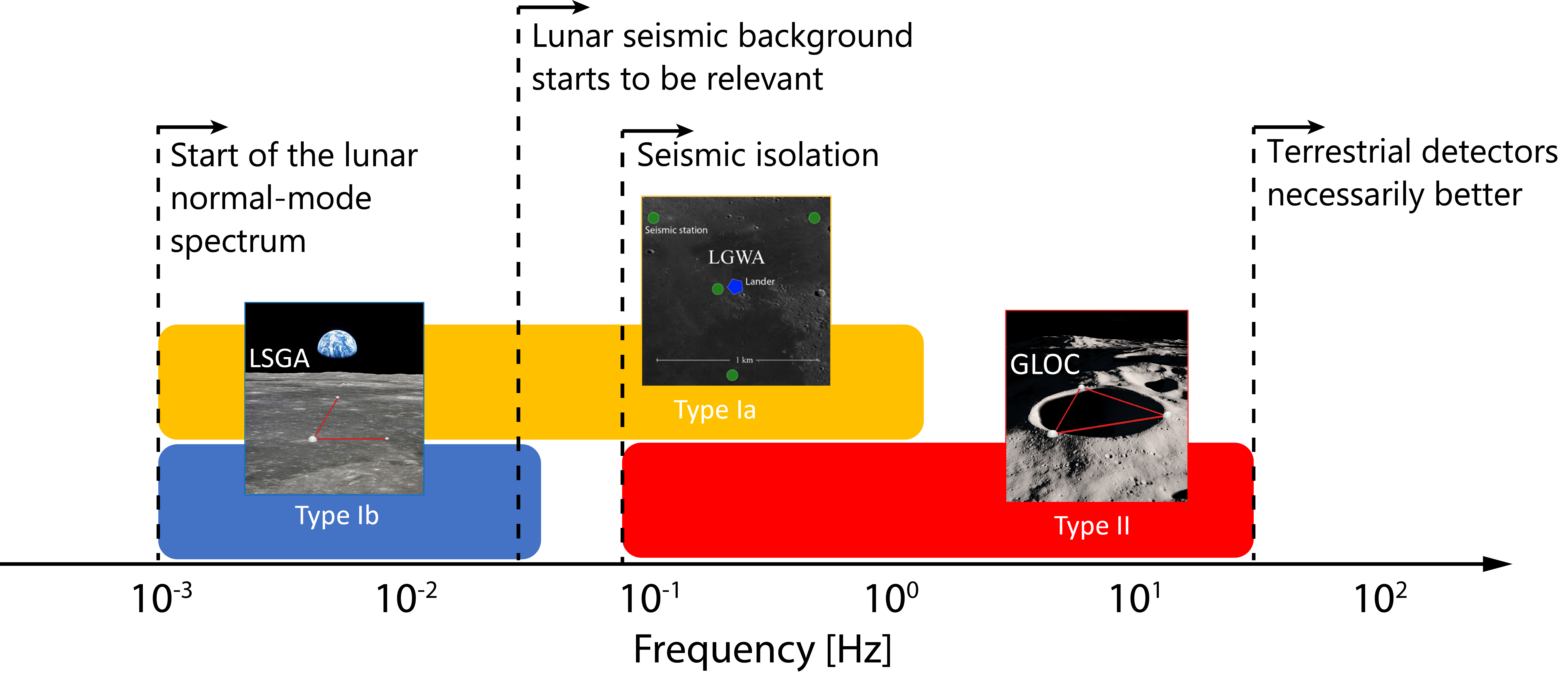}
\caption{Indicative observation bands of recently proposed lunar GW detector concepts based on estimated GW response and seismic background spectrum. The exact boundaries also depend on instrument designs. All concepts still need to undergo feasibility studies.}
\label{fig:concepts}
\end{figure*}
We can now draw conclusions for the conceivable observation bands summarized also in figure \ref{fig:concepts}:
\begin{itemize}[itemindent=2em,leftmargin=2em]
    \item[\tia] An observation band from 1\,mHz to a few Hz is conceivable. The low-frequency limit is set by the frequency of the lowest-order lunar quadrupole mode. The high-frequency end is set by the expected strengths of GW signals compared to the background-limited sensitivity. 
    \item[\tib] An observation band from 1\,mHz to a few 10\,mHz is conceivable. Compared to \tia detectors, the high-frequency bound is lower since \tib detectors experience stronger sensitivity limitations from the seismic background. 
    \item[\tii] The observation band is strongly connected to the detector technologies and weakly constrained by the properties of the lunar seismic background. Therefore, we formulate the bounds as reasonable target. The low-frequency bound should not lie much above 0.1\,Hz since the lunar detector would otherwise lose important advantage over future terrestrial detectors, which have been proposed with observation band down to 3\,Hz \cite{ET2020}. Above a few 10\,Hz, \tii detectors lose their potential advantage over terrestrial detectors since effective isolation and mitigation techniques against environmental disturbances are available.
\end{itemize}

\paragraph{Final considerations} This article touches on two key points of lunar GW detection: the seismic background and the Moon's response to GWs. Both aspects have hot been accurately modeled yet. We only have an upper limit for the seismic background from Apollo data, and its spectrum is basically unknown except for first simplified models. Future seismic missions on the Moon like the approved Farside Seismic Suite \cite{PaEA2021} or the proposed Lunar Geophysical Network \cite{NeEA2020} and LGWA Soundcheck \cite{HaEA2021b} will bring important new insight into the lunar seismic field. 

While the geology of the Moon is not well known \cite{GaEA2011}, enough information is available to set up very useful numerical simulations of the lunar GW response. The challenging part is that the large-scale lunar internal structure as well as kilometer-scale regional structures need to be represented in these models. This might be feasible if the simulation is done for specific deployment sites. 

Finally, we make a brief comment about lunar GW detection below 1\,mHz. At these frequencies, the Moon behaves like a very stiff body, which cannot be significantly deformed by passing GWs. As a consequence, a \tib detector whose entire response relies on the deformability of the Moon is not a good option. Instead, the \tia detector would measure a relatively strong signal $0.5hR_\text{moon}$. However, translating this into a requirement for the sensor's displacement sensitivity and considering current and proposed accelerometer technologies, it seems impossible in the foreseeable future to achieve the required GW strain sensitivity. We therefore exclude the possibility to access this band with lunar GW detectors (however, GW detection \emph{with} the Moon is potentially possible below 1\,mHz \cite{BlJe2022}).

The author thanks Y Chen for useful discussions.

\bibliographystyle{apsrev}
\bibliography{references}

\end{document}